\documentstyle[aps,prl,preprint]{revtex}
\begin{document}
\draft
\preprint{Submitted to PRL}
\title{
Slave Boson Formulation for Interacting Boson Systems and the
Superfluid-Insulator Transition}
\author{Raymond Fr\'esard}
\address {Institut f\"ur Theorie der
Kondensierten Materie, Universit\"at Karlsruhe,\\
76128 Karlsruhe, Germany}
\maketitle
\begin{abstract}
A new formulation for the study of interacting bosons on a lattice
is introduced. This approach is used to give analytical expressions for the
Mott insulating lobes in the phase diagram and to calculate
the density-density correlation function. It is also shown that, at
mean-field level, this newly introduced slave boson theory coincides
with mean-field theory of a suitably introduced order parameter.
\end{abstract}
\newpage
\narrowtext
\baselineskip 0.8cm
\renewcommand{\baselinestretch}{2.0}
\section{INTRODUCTION}
\label{sec:intro}

There has been recently a revival of interest for interaction-induced
metal-to-insulator transitions. This is due to a large extent to the
discovery of high-$T_c$ materials \cite{Bednorz} as in the cuprates
it is believed that
the insulating state originates from a strong Coulomb interaction
in a half-filled band. Anderson's proposal \cite{Anderson} that the
Hubbard Model should capture the essential physics of the cuprates
generated the development and the application of numerous theoretical
techniques to the problem of interacting fermions on the lattice.
In parallel these techniques have been applied to interacting bosons
as well. In a similar way a Mott-insulating state should appear at
commensurate densities due to a strong local interaction that plays the
role of the Pauli principle and prevents the bosons from undergoing
a Bose-Einstein condensation as known from the ideal Bose gas. This
scenario has been recently discussed by Fisher et al \cite{Fisher}
in the framework of a suitable mean-field theory. Their result for
the pure case
has been confirmed by Krauth et al. \cite{Krauth1} using Gutzwiller type
of wave-function, and Sheshadri et al \cite{Sheshadri}. Other mean-field
theories have been proposed \cite{Bruder,Zimanyi}, and Quantum Monte-Carlo
simulations
\cite{Krauth2,Trivedi}, all confirming Fisher et al.'s result. An
alternative technique is provided by introducing auxiliary bosonic
fields. The slave boson representations were pioneered in the context
of spin models by Holstein and Primakoff and, in particular, by Schwinger
who introduced a two slave boson representation of
spin 1/2 \cite{Schwinger}. A second class of slave boson representations
was introduced by Barnes and employed by others, in the context of the
Anderson model of a magnetic impurity in a metal \cite{Barnes}. The method
has been extended to lattice fermion models \cite{Kotliar,Zou,Li89,FW},
and gave rise to an extensive literature. In this letter I introduce a
new representation for both interacting bosons and spins on the lattice
and I apply it to the Bose-Hubbard model. I point out a close connection
between slave boson mean-field theory and the collective
field method. The density-density correlation function in the first
Mott insulating lobe is calculated.

\section{Formulation}
\label{sec:Formu}

The new representation for both interacting
bosons and spins on a lattice is based on a generalization of the
slave boson representation of the fermionic Hubbard Model introduced by
Kotliar and Ruckenstein \cite{Kotliar}, where
each atomic state is obtained with the
help of distinct slave particles subject to constraints. Here the
normalized n-fold
occupied state at a given site $i$ on the lattice is obtained as:
\begin{equation}
\mid n>_i \equiv b^+_{n,i} \mid vac> \hspace{1cm} n\geq 0.
\end{equation}
Even the empty site $\mid 0>_i$ is constructed by operating with the Bose
creation operator $b^+_{0,i}$ onto a new vacuum state $\mid vac>_i$
meaning that even the empty lattice site is not pre-existent but rather
created out of a total vacuum $\mid 0>_i = b^+_{0,i} \mid vac>_i$ in
the enlarged Hilbert space.
The slave bosons $b_{n,i}$ are subject to the constraint:
\begin{equation}
\sum_{n\geq 0} b^+_{n,i} b_{n,i} =1
\end{equation}
indicating that, at each lattice site and at each time, the site has a
well defined occupancy. The eigenvalues of the atomic problem with
on-site interaction $U$
\begin{equation} \label{locei}
E_n=Un(n-1)-\mu n
\end{equation}
serve as chemical
potential for the slave particles. In terms of those the physical boson
creation operator reads:
\begin{equation}
B^+_{i} = \sum_{n\geq 0} \sqrt{1+n} \hspace{3mm} b^+_{n+1,i} b_{n,i}
\end{equation}
It is a straightforward exercise to show that $B$ satisfies the
canonical commutation  relation provided the slave fields do it as well.
This is sufficient in order to rewrite the original Hamiltonian. Taking
as an example the Bose-Hubbard Model:
\begin{equation}
H= \sum_{i,j} t_{i,j}  B^+_{i} B_{j} + \sum_{i} (U(B^+_{i}B_{i}-1) - \mu)
B^+_{i}B_{i}
\end{equation}
The corresponding partition sum as expressed in the slave boson
formalism reads:
\begin{equation}
Z=\int \Pi_{n,i} Db^+_{n+1,i} Db_{n+1,i} \Pi_i D\lambda_i \exp{(-S)}
\end{equation}
with the action
\[S=\int_0^{  \beta} d\tau \sum_{n,i} b^+_{n,i} (\partial_{\tau}
+E_n+i\lambda_i)b_{n,i}  - \sum_i \lambda_i
\]
\begin{equation} \label{act}
+\sum_{i,j,n,m} \sqrt{1+n} \hspace{3mm} \sqrt{1+m}\hspace{3mm}
b^+_{n+1,i} b_{n,i} t_{i,j} b_{m+1,j} b^+_{m,j}
\end{equation}
where the $\lambda$-field enforces the constraint.

The limit of infinite local interaction simply results in restricting
the number of slave fields to be 2. In this case it is worth noting that
the $b$-fields can be taken as fermionic and the action (\ref{act})
describes a quantum $X-Y$ model in an external magnetic
field $\mu/\mu_B$. But in the following I shall keep the $b$-fields as
bosonic. The action (\ref{act}) is invariant under a $U(1)$ gauge
transformation
which allows to eliminate the phase of one slave field
at the price of introducing a time-dependent $\lambda$-field.

\section{Results}
\label{sec:Results}
\noindent a) HARD-CORE LIMIT.

As a first example of handling the action (\ref{act}) I consider the
hard-core limit in the saddle-point approximation. This yields, for
$\mid\mu\mid\leq\mid t_0\mid$, where $t_0 \equiv t(k=0)$:
\begin{equation}
b_0^2 = \frac{t_0+\mu}{2t_0}\hspace{.3cm} , \hspace{.3cm}b_1^2 =
\frac{t_0-\mu}{2t_0} \hspace{.3cm}
and \hspace{.3cm} \lambda_0 = \frac{\mu - t_0}{2}
\end{equation}
and both phase boundaries of the insulating state are properly
recovered. Namely they read $\mu=-2dt$ and $\mu=2dt$ for the empty
$(b_1=0)$ and the full $(b_0=0)$ systems.
Whereas the density $\rho$ is given by
$\mid b_1 \mid^2$, the superfluid density as taken from \cite{Sheshadri}
$\rho_s \sim \mid <B>\mid^2 = \mid b_0b_1\mid^2$
is very different from $\rho$ and they only coincide in the low density
limit. To obtain the gaussian fluctuations
I first take advantage of the $U(1)$
gauge symmetry of $S$ eq. (\ref{act})
in order to take the $b_0$-field as real. I can then integrate out
both $b_0$ and $\lambda$ fields. After having introduced
the mean-field parameters the action reads:
\begin{equation} \label{sgf}
S_{GF} = \frac{1}{2}\sum_k (b'_{1,-k},b''_{1,-k}) \left( \begin{array}{cc}
\alpha_k & \omega_n\\
-\omega_n & \beta_k \end{array} \right) \left( \begin{array}{c}
b'_{1,k}\\b''_{1,k} \end{array} \right)
\end{equation}
where $\alpha_k = b_0^{-2}(t_k-t_0)-4b_1^2t_k$ and $\beta_k =
b_0^2(t_k-t_0)$.
This yields the spectrum
\begin{equation} \label{spectrum}
\omega^2=(t_k-t_0)^2-4b_1^2b_0^2t_k(t_k-t_0)
=_{k\rightarrow 0}2t_0^2b_0^2b_1^2k^2
\end{equation}
which is linear in $k$ for small $k$ as expected in a superfluid state
as a consequence of the spontaneously broken $U(1)$ symmetry.

There is another way to tackle the slave boson action eq. (\ref{act}).
One can decouple the hopping term by introducing a single
Hubbard-Stratonovich field $\Phi$.
The action then becomes bilinear in all
slave boson fields which can then be integrated out exactly. One then
finds
\begin{eqnarray} \label{ww}
Z &=& \int D\Phi D\lambda \exp{\left(-\int_0^{\beta}d\tau (-
\sum_{i,j} \Phi^*_it_{i,j}^{-1}
\Phi_j + \sum_i i\lambda_i)\right)}\nonumber\\
& & \prod_{n,i}
\frac{1}{1-\exp{(-\beta(i\lambda_i + \xi_n)})}
\end{eqnarray}
where, in the occupation number space, ${\xi}_n$ are the eigenvalues of
$M_{m,n} \equiv E_n \delta_{m,n} +
\sqrt{1+m} \Phi \delta_{m+1,n}+ \sqrt{n} \Phi^*\delta_{m,n+1}$,
where the $E_n$ are the eigenvalues of the atomic problem (eq. (\ref{locei}))
and $\Phi$ is treated in mean-field approximation. Due to the
particular form of eq. (\ref{ww}) the constraint can be handled exactly.
In terms of the order parameter $\Phi$, the action reads:
\begin{equation} \label{achs}
S=-\frac{\beta}{t_0}  \mid \Phi \mid^2 - \ln{(\sum_n\exp{(-\beta \xi_n)})}.
\end{equation}
In the hard core limit, the 2 eigenvalues $\xi_{\sigma}$ are given by
$2\xi_{\sigma} = -\mu \pm \sqrt{\mu^2 + 4\mid\Phi\mid^2}$.
This result is well known in the context of the ferromagnetic $X-Y$
model and usually serves as a pedagogical starting point for the
discussion of the paramagnetic-ferromagnetic transition in this model
\cite{Negele}. Solving the saddle-point condition which follows from the
action (\ref{achs}) yields the $(\mu,T)$ phase diagram
where the phase boundary between the superfluid and the normal states
is given by $\mu = 2 T_c th^{-1}(\mu/2dt)$.
This might serve as a description of
the $\lambda$-transition in $^4He$. Assuming that the $^4He$
atoms are sitting on a lattice that is half-filled and that they
only experience a hard-core interaction yields the ratio $T_c/T_{BE}=0.7$,
which is identical to the ratio $T_c^{exp}/T_{BE}$ as obtained with
the help of the experimental data.

\noindent b) FINITE $U$ PROBLEM.

In the saddle-point approximation
the slave-boson action becomes:
\[
S=\sum_{n} \mid b_{n} \mid^2 (E_n+\lambda_0) - \lambda_0 \]
\begin{equation} \label{acmf}
+\sum_{n,m} \sqrt{1+n} \hspace{3mm} \sqrt{1+m}\hspace{3mm}
b^*_{n+1} b_{n} t_{0} b_{m+1} b^*_{m}
\end{equation}

I minimized numerically both eq. (\ref{achs}) and eq. (\ref{acmf}). Even
though both approaches yield very different looking equations, it turns
out that they deliver identical results at zero temperature. The
difficult problem of minimizing the slave-boson action
is handled
in the following way. First of all one phase can be removed owing to the
gauge symmetry of the action. Second the phase of the physical Bose
field can not be determined by the saddle-points equations, as usual in
a superfluid state, but all the others are readily seen to be equal to
the latter, so as to minimize the kinetic energy. There is thus a single
Goldstone mode. The  constraint expresses the fact that the
$N$ bosons to be considered are restricted to the surface of a $N$
dimensional hyper-sphere which I parameterize in polar coordinates. I am
thus left with determining the angles. It turns out that the number of
angles that differ from zero gradually reduces when the interaction is
raised up. This is exemplified in fig. 1 where I show the amplitude of
the first 4 bosonic fields as a function of the interaction strength at
density $\rho=1$. Even for moderate coupling, say $U=t$, only the
first 5 fields differ from 0. This implies that amplitude fluctuations
are very substantially weakened as compared to the ideal Bose gas.
This physical effect persists down to any finite interaction strength.

The Mott insulating state is reached when a single b-field differs from
0. In this case the superfluid density, which vanishes, is very
different from the density which takes (here) the value 1. We meet a
very
different situation as in the weakly interacting Bose gas theory, where
both quantities are identical due to Galilean invariance at zero
temperature \cite{Popov}.
How they start
to deviate from each other is shown in the inset of fig. 1, where I plot
$|<B>|^2/\rho$ as a function of the interaction strength at density
$\rho=1$.

The phase diagram (fig. 2)
shows Mott insulating states corresponding to commensurate
densities, which are identical to those obtained by \cite{Sheshadri}. They
are in good agreement with QMC results by Trivedi and
Ullah \cite{Trivedi}, even though
the insulating lobes are somewhat too small. This is a drawback of the
method which is well-known from similar calculations for interacting
fermions \cite{Kotliar}. This originates in the fact that the action
(\ref{act}), even though exact, does not yield the correct
non-interacting limit at mean-field level.
Phase fluctuations in this approach are somewhat atypical. At finite
temperature they are expected to cause the superfluid-normal transition.
Even though there are a lot of phases that fluctuate, only a single one
is relevant, all other being massive. This is in contrast to the fermionic
Hubbard model where all phases but one can be gauged away and the
fluctuations of the remaining one leads to a massive mode that is then
not expected to destroy the condensate \cite{FW}. Here the fluctuations
bring a rich excitation spectrum.
Considering as an example the gaussian fluctuations in the $n=1$
insulating lobe leads to a decoupling of the propagator matrix. After
having integrated out the (real) $b_1$ field and the constraint field
and introduced $ \alpha_k \equiv E_0 - E_1 +|b_1|^2 t_k$ and
$\beta_k \equiv E_2 - E_1 +2|b_1|^2 t_k$
one obtains:
\[
S_{GF} = \sum_{n\geq 3,k} b^*_{n,k}
(-i \omega_n + E_n -E_1) b_{n,k} \]
\begin{equation}
+ \sum_k (b^*_{0,k},b_{2,-k}) \left( \begin{array}{cc}
-i \omega_n + \alpha_k & \sqrt{2} |b_1|^{2}t_k\\
\sqrt{2} |b_1|^{2}t_k& i \omega_n + \beta_k
\end{array} \right) \left( \begin{array}{c}
b_{0,k}\\b^*_{2,-k} \end{array} \right)
\end{equation}
It follows that the spectrum is split into a set of localized high
energy
excitations (for $n\geq 3$) and a continuum which arises from the small
$n$ part of the action.
Looking for a vanishing gap provides expressions for the
superfluid-insulator lines for the $N$-th lobe ($N \geq 1$):
\begin{equation} \label{moga}
\frac{\mu}{U} = 2N-1 - \frac{dt}{U} \pm
\sqrt{\left(\frac{dt}{U}\right)^2 -  (2N+1) \frac{2dt}{U} + 1}
\end{equation}
This is the analytical expression for the Mott-insulating lobes which
can be obtained either with this method or with the collective field
approach.
In turn physical response functions such as
the density-density correlation function can be computed.
The Mott gap following from the low energy part
vanishes at the tip of the lobe where the spec\-trum is
chan\-ging from
being gapful and massful to gapless and massless. At this
particular point I calculated numerically for $N=1$
the den\-si\-ty-den\-si\-ty
cor\-re\-la\-tion function $N(q,\omega)$
on the 2-d square lattice which is displayed in
fig. 3. The latter is vanishing iden\-ti\-cal\-ly for $k=0$ reflecting
the in\-com\-pres\-si\-bi\-li\-ty of the system. As a result
charge fluctuations are
mostly high energetic and inhomogeneous up to a
fraction of low but finite energy excitations resulting into a gapless
incompressible state. The spectrum is very broad and is extending well
over the band width of the corresponding non-interacting Fermi system.
It mostly consists of a two-peak structure following from the two modes
of the fluctuation matrix.
At low momentum and energy the imaginary part of $N(q,\omega)$ is
obtained as:
\begin{equation} \label{imn}
Im N(q,\omega) = \frac{\Theta(\omega -cq)}{16}
\frac{q^2}{\sqrt{\omega^2 -(cq)^2}}
\end{equation}
with the sound velocity $c=4t/\sqrt{6\sqrt{2}-8}$. This response
function thus exhibits an integrable singularity at the threshhold.
As a result the DC
conductivity is finite and takes the universal value $e^2/16 \hbar$
where $e$ denotes the charge of the bosons. However a finite value for
the DC conductivity heavily relies on the particular form of $Im
N(q,\omega)$ as found in eq (\ref{imn}). Keeping in mind that all
self-energy corrections to the slave boson propagators are neglected
makes it unlikely that such a peaked behavior is a genuine feature of
the model. Moreover an additional symmetry of the Hamiltonian
appearing at a discrete set of points of the phase diagram
corresponding to the tip of the lobes could not be identified. Thus
self-energy corrections must exist. Here the holon and doublon
propagators only coincide for small energy and momentum. The
differences are responsible for the 2-peak structure in $Im
N(q,\omega)$ rather than a 1-peak structure.
Calculations of $N(q,\omega)$
away from the tip of the lobe yields similar looking results apart from a
gap as obtained from eq (\ref{moga}). The detailed expression
for $N(q,\omega)$
as well as additional calculations in the superfluid domain that are
in progress, will be published elsewhere \cite{laf}.

\centerline{\bf Conclusion}

In this paper the superfluid-insulator transition that occurs
in interacting bosons systems is considered. To this
aim  an auxiliary boson representation is introduced
and I showed that the slave boson mean-field theory is
identical to a mean-field theory on an order parameter. Despite of its
apparent complexity it allowed to obtain
an analytical expression for the Mott insulating
lobes  as well as for the density-density correlation
function. This newly introduced framework is used to show that many
energy scales appear in the excitation spectrum. In the strong
coupling regime it consists of a continuum of low energy excitations
and a set of localized high energy excitations. The latter are
weakening the amplitude fluctuations in a substantial way, even for
moderate coupling.
\acknowledgments

It is a pleasure to thank Prof. P. W\"olfle for many enlightening
discussions
and Dr. Christoph Bruder as well as Dr. A. van Otterlo for interesting
discussions.
I thank the Deutsche Forschungsgemeinschaft for financial support under
Sonderforschungsbereich 195.

\newpage
FIGURE CAPTIONS.
FIG. 1 Amplitude of the slave boson fields $b_n$ as functions of the
interaction strength at the commensurate density $\rho=1$. The curves
A,B,C,D correspond respectively to $n=0,1,2,3$. Inset: $|<B>|^2$
as a function of the interaction strength
at the commensurate density $\rho=1$.

FIG. 2 Chemical potential versus hopping phase diagram at zero
temperature. The insulating lobes correspond to the densities
1 and 2.

FIG. 3 Density-density correlation function at the tip of the first
Mott lobe on the square lattice as functions of frequency. The curves
A, B, C, D
are calculated for wave-vectors on the diagonal of the first Brillouin
zone for respectively $q_x= (1, 2, 3, 5) \pi/5$.

\end{document}